\documentclass[11pt]{article}

\usepackage{etoolbox}
\newtoggle{submission}
\togglefalse{submission}

\usepackage[utf8]{inputenc}
\usepackage[T1]{fontenc}
\usepackage{courier}

\usepackage{xcolor}
\definecolor{bpforms}{cmyk}{1,0,0,0}
\definecolor{mod}{rgb}{1.0, 0.55, 0.0}
\definecolor{attr}{rgb}{0.52, 0.73, 0.4}
\definecolor{val}{rgb}{1.0, 0.44, 0.37}
\definecolor{lightgrey}{rgb}{0.925, 0.925, 0.925}

\usepackage[breaklinks, colorlinks=true, citecolor=bpforms, linkcolor=attr, urlcolor=mod]{hyperref}

\usepackage{graphicx}
\DeclareGraphicsExtensions{.pdf,.PDF}

\usepackage{enumitem}
\setitemize{topsep=0pt, partopsep=0pt}
\usepackage[margin=1in]{geometry}

\usepackage[compact,explicit]{titlesec}

\titleformat{\section}[block]{\bfseries\Large}{\thesection.}{1ex}{#1}[]
\titlespacing*{\section}{0pt}{*2.0}{*0.0}

\titleformat{\subsection}[block]{\bfseries\large}{\thesubsection.}{1ex}{#1}[]
\titlespacing*{\subsection}{0pt}{*0.0}{*-1.0}

\titleformat{\subsubsection}[runin]{\bfseries}{}{0ex}{#1.}[]
\titlespacing*{\subsubsection}{0pt}{*0.0}{1ex}

\titleformat{\paragraph}[runin]{\itshape}{}{0em}{#1.}[]
\titlespacing*{\paragraph}{0pt}{*0.0}{1ex}

\usepackage{setspace}

\makeatletter
\newcommand\Autoref[1]{\@first@ref#1,@}
\def\@throw@dot#1.#2@{#1}
\def\@set@refname#1{
    \edef\@tmp{\getrefbykeydefault{#1}{anchor}{}}%
    \xdef\@tmp{\expandafter\@throw@dot\@tmp.@}%
    \ltx@IfUndefined{\@tmp autorefnameplural}%
         {\def\@refname{\@nameuse{\@tmp autorefname}s}}%
         {\def\@refname{\@nameuse{\@tmp autorefnameplural}}}%
}
\def\@first@ref#1,#2{%
  \ifx#2@\autoref{#1}\let\@nextref\@gobble
  \else%
    \@set@refname{#1}
    \@refname~\ref{#1}
    \let\@nextref\@next@ref
  \fi%
  \@nextref#2%
}
\def\@next@ref#1,#2{%
   \ifx#2@ and~\ref{#1}\let\@nextref\@gobble
   \else, \ref{#1}
   \fi%
   \@nextref#2%
}
\makeatother

\usepackage[
    backend=bibtex,
    style=nature,
    citestyle=numeric-comp,
    sorting=none,
    natbib=true,
    autocite=superscript,
    maxbibnames=10
    ]{biblatex}
\let\cite=\autocite
\newcommand{\dontcite}[1]{}

\usepackage{newfloat}
\DeclareFloatingEnvironment[fileext=frm,placement={!ht},name=Box,within=none]{floatbox}
\usepackage[labelfont=bf,singlelinecheck=false,labelsep=period, skip=6pt]{caption}
\iftoggle{submission}{}{
}

\usepackage{tcolorbox}
\tcbset{colback=lightgrey, sharp corners=all, frame empty, 
right=0em, left=0em, top=0em, bottom=0em, before skip=2pt, after skip=10pt}

\newlength{\boxwidth}
\usepackage{placeins}

\iftoggle{submission}{
    \usepackage{lineno}
}{}

\newcommand{\BpForms}{\textit{BpForms}}
\newcommand{\BcForms}{\textit{BcForms}}

\usepackage{authblk}

\title{\vspace{-2.6em}{\BpForms} and {\BcForms}: Tools for concretely describing non-canonical polymers and complexes to facilitate comprehensive biochemical networks}

\author[1,2,3,*]{Paul F. Lang}
\author[1,2,4,*]{Yassmine Chebaro}
\author[1,2,*]{Xiaoyue Zheng}
\author[1,2]{John A. P. Sekar}
\author[1,2]{Bilal Shaikh}
\author[5]{Darren A. Natale}
\author[1,2,**]{Jonathan R. Karr}
\affil[1]{Icahn Institute for Data Science and Genomic Technology, Icahn School of Medicine at Mount Sinai, New York, NY 10029, USA}
\affil[2]{Department of Genetics and Genomic Sciences, Icahn School of Medicine at Mount Sinai, New York, NY 10029, USA}
\affil[3]{Department of Biochemistry, Oxford University, South Parks Road, Oxford OX1 3QU, UK}
\affil[4]{Institut de G\'en\'etique et de Biologie Mol\'eculaire et Cellulaire, Institut National de la Sant\'e et de la Recherche M\'edicale, Centre National de la Recherche Scientifique, Universit\'e de Strasbourg, 67404, Illkirch, France}
\affil[5]{Protein Information Resource, Georgetown University Medical Center, Washington, DC 20007, USA\authorcr~}

\affil[*]{These authors contributed equally to this work}
\affil[**]{Correspondence: \href{mailto:karr@mssm.edu}{karr@mssm.edu}\vskip-1em}

\date{August 26, 2019}

\begin{document}
\maketitle
\iftoggle{submission}{
    \linenumbers
    \doublespacing
}{}

\section*{Abstract}

Although non-canonical residues, caps, crosslinks, and nicks play an important role in the function of many DNA, RNA, proteins, and complexes, we do not fully understand how networks of non-canonical macromolecules generate behavior. One barrier is our limited formats, such as IUPAC, for abstractly describing macromolecules. To overcome this barrier, we developed {\BpForms} and {\BcForms}, a toolkit of ontologies, grammars, and software for abstracting the primary structure of polymers and complexes as combinations of residues, caps, crosslinks, and nicks. The toolkit can help quality control, exchange, and integrate information about the primary structure of macromolecules into fine-grained global networks of intracellular biochemistry.

\section*{Keywords}
format; software; polymer; proteoform; complex; residue; modification; crosslink; fine-grained network; genome-scale network

\section{Background}

A central goal in biology is to understand how networks of metabolites, DNA, RNA, proteins, and complexes generate behavior. Non-canonical residues, caps, crosslinks, and nicks are essential to these networks. For example, prokaryotic restriction/modification systems use methylation to selectively degrade foreign DNA, tRNA use pseudouridine to translate multiple codons, and signaling networks use phosphorylation to encode information into the states of proteins. 

Recent technical advances have enabled detailed information about individual DNA, RNA, and protein modifications. For example, SMRT-seq can identify the locations of DNA methylations with single-nucleotide resolution \cite{plongthongkum2014advances} and mass-spectrometry can identify hundreds of protein modifications \cite{toby2016progress}. Furthermore, several repositories have compiled extensive data about non-canonical residues and crosslinks in DNA \cite{sood2019dnamod, milanowska2010repairtoire, ye2016methsmrt, xuan2017rmbase}, RNA \cite{boccaletto2017modomics, cantara2010rna}, and proteins \cite{montecchi2008psi, garavelli2004resid, hornbeck201815, rose2012rcsb, natale2016protein, uniprot2018uniprot}, as well data about the subunit composition and crosslinks of complexes \cite{rose2012rcsb, uniprot2018uniprot, meldal2018complex, giurgiu2018corum, karp2017biocyc}. Despite this progress, it remains difficult to integrate this information into fine-grained global networks of intracellular biochemistry, in part, because these resources use chemically-ambiguous and incompatible formats. Consequently, we still do not have a holistic understanding of how non-canonical macromolecules help generate behavior.

Whole-cell (WC) models \cite{karr2012whole, goldberg2018emerging}, which aim to predict phenotype from genotype by representing all of the biochemical activity in cells, are a promising tool for integrating diverse information about macromolecules into a holistic understanding of cellular behavior. However, it remains challenging to build fine-grained, global biochemical networks, such as WC models, because we have few tools for capturing the structures of non-canonical macromolecules and linking them together into networks. For example, formats such as BioNetGen \cite{harris2016bionetgen} and the Systems Biology Markup Language (SBML) \cite{hucka2018systems} are cumbersome for modeling post-transcriptional modification because they have limited capabilities to represent the primary structure of RNA \cite{misirli2015annotation, courtot2011controlled}. Abstractions of the primary structures of macromolecules that can be combined with modeling frameworks such as SBML would provide a significant step toward fine-grained global biochemical networks. Combined with software tools, such abstractions could also facilitate the curation, exchange, and quality control of structural information about macromolecules for a wide range of omics and systems and synthetic biology research.

Currently, several formats have limited abilities to abstract the primary structures of non-canonical polymers and complexes. Molecular formats which represent each atom and bond, such as the International Chemical Identifier (InChI) \cite{heller2015inchi}, the PDB format \cite{westbrook2003pdb}, and the Simplified Molecular-Input Line-Entry System (SMILES) \cite{weininger1988smiles}, can represent non-canonical residues, caps, crosslinks, and nicks. However, their fine granularity is cumbersome for network-scale research. Omics and systems biology formats, such as BioPAX \cite{demir2010biopax}, the Biological Expression Language (BEL) \cite{fluck2016training}, the MODOMICS nomenclature \cite{boccaletto2017modomics}, the PRO notation \cite{natale2016protein}, ProForma \cite{leduc2018proforma}, and the Synthetic Biology Open Language (SBOL) \cite{cox2018synthetic}, use abstractions that are conducive to network-scale research. However, these formats have limited abilities to represent non-canonical residues, caps, crosslinks and nicks, and they do not concretely represent the primary structures of macromolecules. 

Toward fine-grained global networks of intracellular biochemistry, we developed {\BpForms}-{\BcForms}, an open-source toolkit for abstractly representing the primary structure of polymers and complexes. {\BpForms} includes extensible alphabets of hundreds of DNA, RNA and protein residues; an ontology of common crosslinks; and a human and machine-readable grammar for combining residues, residue modifications, intra-chain crosslinks, and nicks into polymers. {\BcForms} includes a human and machine-readable grammar for combining polymers, small molecules, and inter-chain crosslinks into complexes. Both tools include software for validating descriptions of macromolecules, calculating properties of macromolecules such as their formula, visualizing macromolecules, and exporting macromolecules to molecular formats such as SMILES. Both tools are available as a web application, REST API, command-line program, and Python library.

Here, we describe the toolkit and demonstrate how it can facilitate omics, systems modeling, and synthetic biology. First, we describe the toolkit, including the alphabets of residues, the ontology of crosslinks, the grammars, the software tools, and the user interfaces. Second, we describe how {\BpForms} and {\BcForms} can be integrated with knowledge about pathways, kinetic models, and genetic designs through formats such as BioPAX, CellML \cite{cuellar2015cellml}, SBML, and SBOL. Next, we describe the advantages of the toolkit over existing formats for representing polymers and complexes and existing alphabets of residues. Lastly, we present multiple case studies that illustrate how the toolkit can help researchers describe, quality control, exchange, and integrate diverse information about macromolecules into networks. We anticipate that {\BpForms} and {\BcForms} will help facilitate fine-grained, global networks of cellular biochemistry.

\section{Results}

\subsection{Toolkit for abstracting non-canonical polymers and complexes}
\begin{figure}[!ht]
\centering
\iftoggle{submission}{}{%
    \includegraphics[width=\textwidth]{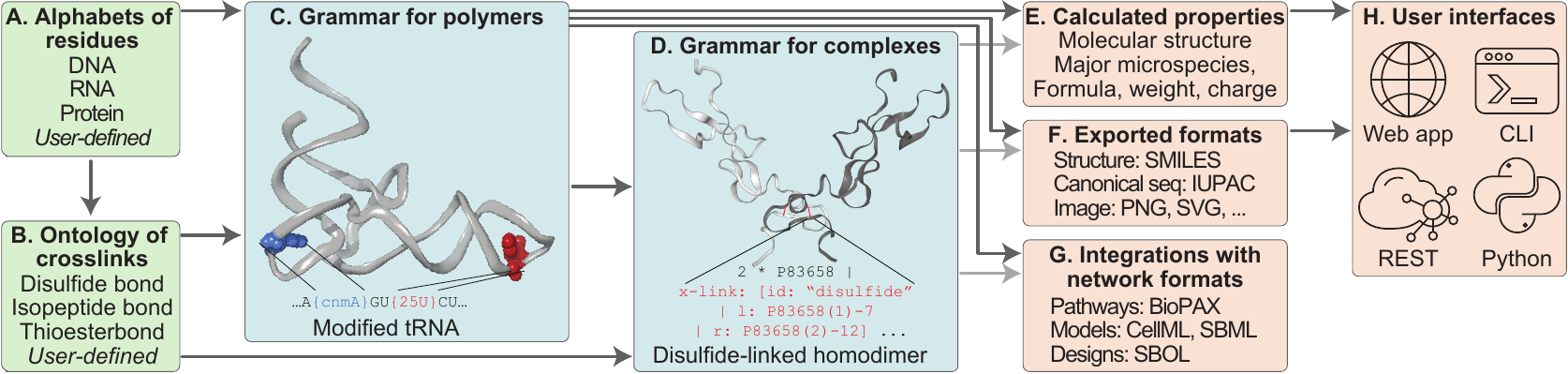}%
}%
\caption{\label{fig:overview} \textbf{The {\BpForms}-{\BcForms} toolkit can abstract, validate, and analyze the primary structures of non-canonical polymers and complexes and help integrate structural information about macromolecules into networks.} The toolkit includes (\textbf{A}) extensible alphabets that represent individual DNA, RNA and protein residues; (\textbf{B}) an ontology of crosslinks; (\textbf{C}) a grammar for composing polymers from residues, caps, crosslinks and nicks; (\textbf{D}) a grammar for composing complexes from polymers and crosslinks; software tools for validating descriptions of macromolecules, (\textbf{E}) calculating molecular properties of macromolecules, (\textbf{F}) exporting macromolecules to other formats, and visualizing macromolecules; (\textbf{G}) protocols for integrating structural information about macromolecules into omics, systems biology, and synthetic biology formats for networks, models, and genetic designs; and (\textbf{H}) multiple user interfaces.}
\end{figure}

The {\BpForms}-{\BcForms} toolkit includes several interrelated tools for describing, validating, visualizing, and calculating properties of the primary structure of DNA, RNA, proteins, and complexes (\autoref{fig:overview}). Here, we describe the components of the toolkit including the abstractions and grammars for polymers and complexes; the alphabets of residues; the ontology of crosslinks; the software tools for quality controlling, analyzing, and visualizing macromolecules; the protocols for integrating {\BpForms} and {\BcForms} with formats for network research; and the user interfaces.

\subsubsection{Abstract representation of the primary structure of polymers and complexes}
\begin{figure}[!ht]
\centering
\iftoggle{submission}{}{%
    \includegraphics{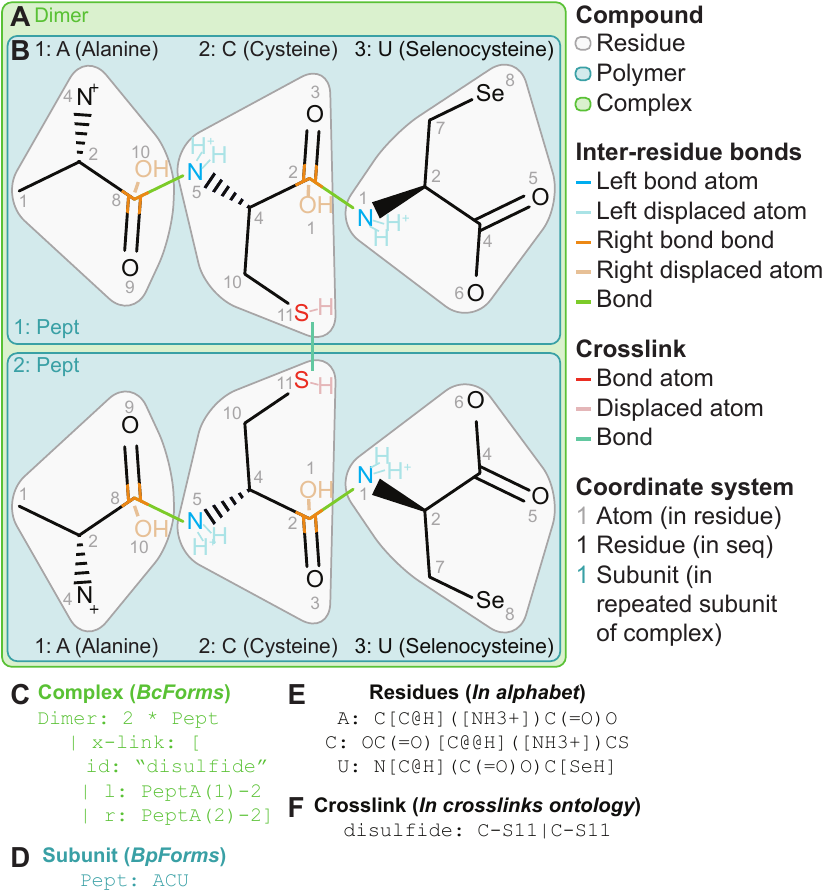}%
}
\caption{\label{fig:representation} \textbf{{\BpForms} and {\BcForms} abstract the primary structures of polymers and complexes as combinations of residues, crosslinks, and nicks.} For example, {\BcForms} abstracts a disulfide-linked homodimer (\textbf{A}, green box) of a selenocysteine-modified tripeptide (\textbf{B}, blue boxes) as two copies of the tripeptide and a single crosslink (\textbf{C}, green text) and {\BpForms} abstracts the peptide as a sequence of three residues, including selenocysteine (U) (\textbf{D}, blue text). These abstractions are enabled by alphabets of residues (\textbf{E}, black text) and an ontology of crosslinks (\textbf{F}, black text). 
}
\end{figure}

{\BpForms} represents polymers as a sequence of residues, a set of crosslinks, a set of nicks, and a Boolean indicator of circularity (\autoref{fig:representation}B, D). {\BcForms} represents complexes as a set of subunits and a set of crosslinks (\autoref{fig:representation}A, C). Each subunit is represented by its molecular structure and stoichiometry. The structure of each subunit can be described using {\BpForms} or SMILES.

\paragraph{Residues}
Each residue is represented by its molecular structure, a list of the atoms which can form bonds with preceding and following residues, and a list of the atoms which are displaced by the formation of these bonds (\autoref{fig:representation}E). These lists of atoms are optional to enable the toolkit to represent internal nucleic and amino acids, as well as 3' and 5' caps. The toolkit can also capture metadata and missing information about residues. 

\paragraph{Crosslinks}
Each crosslink is represented as lists of the atoms which can form a bond between residues and the atoms which are displaced by the formation of these bonds (\autoref{fig:representation}F). The toolkit represents each nick as a tuple of adjacent residues which are not bonded.

\paragraph{Alphabets of residues and ontology of crosslinks}
The toolkit uses a hybrid approach to abstract the molecular details of residues and crosslinks from the descriptions of macromolecules. The chemical details of common residues and crosslinks are abstracted into alphabets of residues and an ontology of crosslinks. Users can define additional residues and crosslinks within descriptions of macromolecules or create custom alphabets and ontologies. This hybrid approach standardizes the representation of common residues and crosslinks while enabling the toolkit to represent any residue or crosslink.

\paragraph{Coordinate system}
The toolkit uses a structured coordinate system to describe the atoms involved in each inter-residue bond and crosslink. The coordinate of each repeated subunit ranges from one to the stoichiometry of the subunit. The coordinate of each residue is its position within the residue sequence of its parent polymer. The coordinate of each atom is its position within the canonical SMILES ordering of the atoms in its parent residue. \textcolor{attr}{Additional File~1.4} contains more information about the coordinate system.

\paragraph{Examples}
\Autoref{box:bpforms-grammar, box:bcforms-grammar} illustrate the toolkit's grammars for describing polymers and complexes, and \autoref{fig:representation} illustrates the chemical semantics of a homodimer encoded in the grammars. \textcolor{attr}{Additional File~1.2} and the {\BpForms} and {\BcForms} websites provide detailed descriptions of the grammars and additional examples. \textcolor{attr}{Additional File~1.3} contains formal descriptions of the grammars.

\subsubsection{Alphabets of DNA, RNA, and protein residues}
To support a broad range of research, {\BpForms} includes the most extensive alphabets of DNA, RNA, and protein residues to date. The DNA alphabet includes 422 deoxyribose nucleotide monophosphates and 3' and 5' caps derived from data about DNA damage and repair from REPAIREtoire \cite{milanowska2010repairtoire}, structural data from the Protein Data Bank Chemical Component Dictionary (PDB CCD) \cite{westbrook2014chemical}, and chemoinformatics data from DNAmod \cite{sood2019dnamod}. The RNA alphabet includes 378 ribose nucleotide monophosphates and 3' and 5' caps derived from biochemical data from MODOMICS \cite{machnicka2012modomics} and the RNA Modification Database \cite{cantara2010rna} and structural data from the PDB CCD. The protein alphabet has 1,435 amino acids and carboxy and amino termini derived from biochemical data from RESID \cite{garavelli2004resid} and structural data from the PDB CCD. The {\BpForms} website contains pages which display the residues in each alphabet. \textcolor{attr}{Additional File~1.5} describes how we constructed the alphabets.

\subsubsection{Ontology of crosslinks}
To abstract the molecular structures of polymers and complexes, the toolkit includes the first ontology of crosslinks. Currently, the ontology contains 36 common crosslinks. We plan to continue to curate additional crosslinks as needed to represent WC models. The {\BpForms} website contains a page which displays the crosslinks in the ontology. \textcolor{attr}{Additional File~1.6} describes how we constructed the ontology.

\subsubsection{Syntactic and semantic validation of descriptions of macromolecules}
To help quality control information about macromolecules, the toolkit can verify the syntactic and semantic correctness of macromolecules encoded in {\BpForms} and {\BcForms}. First, the toolkit can verify that textual descriptions of macromolecules are syntactically consistent with the {\BpForms} and {\BcForms} grammars and identify any errors. Second, the toolkit can verify that macromolecules represented by {\BpForms} and {\BcForms} are semantically consistent and identify any errors. For example, the toolkit can identify pairs of adjacent amino acids that cannot form peptide bonds because the first amino acid does not have a carboxy terminus or because the second amino acid does not have an amino terminus. \textcolor{attr}{Additional File~1.7} details the semantic validations implemented by the toolkit. We anticipate that these quality controls will help researchers exchange reliable information and assemble this information into high-quality networks.

\subsubsection{Analyses of polymers and complexes}
The toolkit can calculate several properties of macromolecules such as their primary structure, major protonation and tautomerization states, chemical formula, molecular weight, and charge. We have begun to use these properties to quality control WC models. For example, we are using the chemical formulae to verify that each reaction is element and charge balanced, including reactions that represent transformations of macromolecules such as the post-transcriptional modification of tRNA. 

The toolkit can also compare macromolecules to determine their equality or identify differences. We plan to use this feature to implement automated procedures for merging models that share species and reactions.

\subsubsection{Molecular and sequence visualizations}
To help analyze macromolecules, the toolkit can generate molecular and sequence visualizations of residues, caps, crosslinks, polymers, and complexes. The molecular visualizations display each atom and bond and use colors to highlight features such as individual residues, inter-residue and crosslink bonds, and the atoms that are displaced by the formation of the inter-residue bonds (\textcolor{attr}{Figure~S1A--C}). The molecular visualizations can also display the coordinate of each residue and atom. The sequence visualizations include interactive tooltips that describe each non-canonical residue, crosslink, and nick (\textcolor{attr}{Figure~S1D}).

\subsubsection{Export to other molecular and sequence formats}
For compatibility with structural and biochemical research, the toolkit can export {\BpForms} and {\BcForms}-encoded macromolecules to molecular formats such as InChI, the PDB format, and SMILES. For compatibility with genomics research, the toolkit can also export the canonical sequences of {\BcForms}-encoded polymers to the IUPAC/IUBMB format \cite{leonard2003iupac} and FASTA documents \cite{pearson1990rapid}.

\subsubsection{Integration with frameworks for network-scale research}
{\BpForms} and {\BcForms} can facilitate network-scale research through integration with omics and systems and synthetic biology frameworks such as BioPAX, CellML, SBML, and SBOL. \textcolor{attr}{Additional File~1.9} illustrates how {\BpForms} and {\BcForms} can be incorporated into these frameworks.

\subsubsection{User interfaces}
{\BpForms} and {\BcForms} each include four user-friendly interfaces: a web application, a REST API, a command-line program, and a Python library.

\subsection{\label{sec:comparison} Comparison with existing formats and alphabet-like resources}
{\BpForms} and {\BcForms} are the first abstractions that can represent the primary structure of any DNA, RNA, protein, and complex, including non-canonical residues, caps, crosslinks, nicks, and circularity. The toolkit also contains the most extensive alphabets of DNA, RNA, and protein residues and the first ontology of concrete crosslinks. Furthermore, the toolkit has several innovative features to facilitate research about non-canonical macromolecules: the toolkit includes a novel coordinate system that makes it easy to address specific atoms in macromolecules, the toolkit uses a novel combination of ontologies and inline definitions of residues and crosslinks to standardize the representation of common residues and crosslinks while accommodating any residue or crosslink, and the toolkit includes novel quality controls for abstractions of the primary structures of macromolecules. Taken together, {\BpForms} and {\BcForms} are well-suited for network research. Here, we summarize how {\BpForms} and {\BcForms} improve upon several existing resources for abstracting polymers and complexes.

\subsubsection{Comparison of {\BpForms} with existing formats for polymers}
{\BpForms} is the first format that can abstract the primary structure of DNA, RNA, and proteins, including non-canonical residues, caps, crosslinks, nicks, and circularity. In contrast, molecular formats such as SMILES do not abstract the structures of polymers, and abstract formats such as ProForma and network formats such as BioPAX do not represent concrete molecular structures. {\BpForms} also provides a unique blend of the features of previous molecular and abstract formats: {\BpForms} can capture missing information similar to ProForma, {\BpForms} is human-readable like other abstract formats, {\BpForms} is machine-readable like molecular formats, {\BpForms} is composable with network formats such as SBML like molecular formats, and {\BpForms} is backward compatible with the IUPAC/IUBMB format like other abstract formats. \textcolor{attr}{Additional File~1.11.1} and \textcolor{attr}{Table~S1} provide a detailed comparison of {\BpForms} with several other formats.

\subsubsection{Comparison of {\BpForms} alphabets with existing databases}
The {\BpForms} alphabets are the most extensive alphabets of DNA, RNA, and protein residues because they are based on structural, biochemical, and physiological data from several sources. In addition, the {\BpForms} alphabets and the PDB CCD are the only alphabets which consistently represent DNA, RNA, and protein residues and which represent the inter-residue bonding sites of each residue, enabling residues to be combined into concrete molecular structures. In contrast, DNAmod, REPAIRtoire, MODOMICS, RESID, and the RNA Modification Database each only represent DNA, RNA, or protein residues; the residues in DNAmod, REPAIRtoire, MODOMICS, and the RNA Modification Database are hard to compose into polymers because they represent nucleobases and nucleosides rather than nucleotides; and DNAmod, REPAIRtoire, MODOMICS, RESID, and the RNA Modification Database do not capture bonding sites. \textcolor{attr}{Additional File~1.11.2} and \textcolor{attr}{Table~S2} provide a detailed comparison of the {\BpForms} alphabets with several other resources.

\subsubsection{Comparison of the {\BpForms} crosslinks ontology with existing resources}
Several resources contain information about crosslinks. In particular, the UniProt controlled vocabulary of posttranslational modifications includes textual descriptions of over 100 types of crosslinks. In addition, MOD, REPAIRtoire, and RESID indirectly represent crosslinks by representing crosslinked dimers and trimers. 

The {\BpForms} ontology is the first resource which directly represents the chemical structures of crosslinks, enabling crosslinks to be composed into concrete structures. In contrast, MOD, REPAIRtoire, and RESID represent crosslinks indirectly and the crosslinks in UniProt do not have concrete chemical semantics. Consequently, the crosslinks in MOD, REPAIRtoire, RESID, and UniProt cannot be composed into concrete structures. \textcolor{attr}{Additional File~1.11.3} and \textcolor{attr}{Table~S3} provide a detailed comparison of the {\BpForms} crosslinks ontology with these resources.

\subsubsection{Comparison of {\BcForms} with existing formats for complexes}
Despite the importance of complexes, only a few formats can represent complexes. The PDB format is well-suited to capturing the 3-dimensional structures of complexes. BioPAX and SBOL can also capture the subunit composition of complexes. 

{\BcForms} is the first format which abstracts the primary structures of complexes including crosslinks. In contrast, the PDB format has limited capabilities to abstract crosslinks, and BioPAX and SBOL have limited abilities to represent stochiometric information and crosslinks. {\BcForms} is also the first format which can be composed with formats for networks such as SBML. \textcolor{attr}{Additional File~1.11.4} and \textcolor{attr}{Table~S4} provide a detailed comparison of {\BcForms} with several other formats.

\subsection{Case studies}
We believe that the {\BpForms}-{\BcForms} toolkit can support a wide range of omics and systems and synthetic biology research. Here, we illustrate how we have used the toolkit to improve the quality of the PRO database of proteoforms; analyze the metabolic cost of tRNA modification in \textit{Escherichia coli}; refine, expand, a compose a model of MAPK signaling with models of other pathways; and identify constraints on designing new strains of \textit{E. coli}.

\subsubsection{Proteomics: Quality control of the Protein Ontology}
One of the goals of proteomics is to characterize the proteoforms in cells. Toward a comprehensive catalog of proteoforms, the PRO consortium has manually integrated several different types of data into PRO, a database of 8,095 proteoforms. Because the consortium constructs PRO, in part, by hand, automated quality controls could help the consortium identify and correct errors in PRO. 

We have used {\BpForms} quality control PRO. First, we encoded each entry in PRO into the {\BpForms} grammar and used the {\BpForms} software to validate each entry. This identified several types of syntactical and semantic errors. For example, we identified annotated processing sites that have invalid coordinates that are greater than the length of the translated sequence of their parent protein. We also identified modified residues whose structures are inconsistent with the translated sequences of their parent proteins, such as a phosphorylated serine which is annotated at the position of a tyrosine in the translated sequence of its parent. Second, the consortium corrected these errors. These improvements will be published with the next release later this year.

To enable the consortium to continue to use {\BpForms} to quality control PRO, we developed a script which automates this analysis. Going forward, the consortium also plans to use {\BpForms} and {\BcForms} to visualize and export proteoforms to molecular formats such as SMILES.

\subsubsection{Systems biology: Analysis of the metabolic cost of prokaryotic tRNA modification}
To achieve WC models, we must integrate information about all of the processes in cells and their interactions. Here, we illustrate how {\BpForms} can help integrate information about the interaction between the RNA modification and metabolism of \textit{E. coli} and identify gaps in models.

\begin{figure}[!ht]
\centering
\iftoggle{submission}{}{%
    \includegraphics{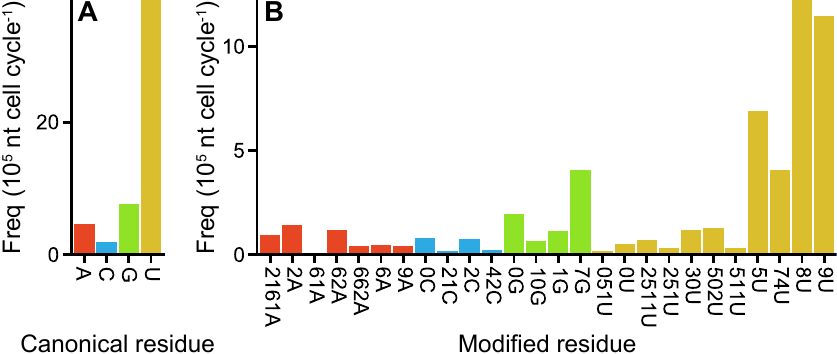}%
}%
\caption{\label{fig:modomics} \textbf{{\BpForms} and {\BcForms} can facilitate integrative analyses of fine-grained global intracellular networks.} For example, we used {\BpForms} to estimate the metabolic cost of tRNA modification in \textit{E. coli} by canonical residue (\textbf{A}) and modified residue (\textbf{B}) from information about the modification, abundance, and turnover of each tRNA.}
\end{figure}

First, we estimated the abundance of each tRNA from the total observed abundance of tRNA \cite{dong1996co, mackie2013rnase} and the observed relative abundance of each tRNA \cite{wei2019improved}. Second, we estimated the synthesis rate of each tRNA from the estimated abundance of each tRNA, the observed half-life of tRNA\textsuperscript{Asn} \cite{bailly2006single}, and the observed doubling time of \textit{E. coli} in glucose media \cite{woldringh1977morphological}. Third, we used {\BpForms} to analyze the curated modifications of each tRNA \cite{boccaletto2017modomics}. Fourth, we estimated the total synthesis rate of each modification from the synthesis rate and modification of each tRNA (\autoref{fig:modomics}). 

This analysis revealed that \textit{E. coli} tRNA contain 26 modified residues, and that the five most abundant residues account for 73.8\% of all modifications. Next, we tried to use the iML1515 metabolic model \cite{monk2017iml1515}, one of the most comprehensive models of cellular metabolism, to analyze the impact of these modifications on metabolism and understand how \textit{E. coli} allocates its limited metabolic resources among these modifications. This analysis revealed that the model only represents one of the modified residues (9U, pseudouridine). Therefore, the model must be expanded to capture the metabolic cost of tRNA modification.

\subsubsection{Systems biology: Systematic identification of gaps in the Kholodenko model of MAPK signaling}
The Kholodenko model of the eukaryotic MAPK signaling cascade \cite{kholodenko2000negative} describes how the cascade transduces extracellular signals for growth, differentiation, and survival into the phosphorylation state of MAPK. However, the model does not account for factors such as the cell's nutritional status. 

Toward a more holistic model of the cascade, we used {\BpForms} to systematically identify gaps in the Kholodenko model and opportunities to merge the model with models of other pathways. First, we obtained an SBML-encoded version of the model. Second, we determined the specific proteins represented by the model. We had to do this manually because Kholodenko did not report this information. Third, we curated the sequences and post-translational modifications of the species represented by the model from UniProt and encoded them into {\BpForms} (\autoref{fig:kholodenko}A). Fourth, we embedded these {\BpForms} representations into the SBML representation of the model. We believe that the {\BpForms} annotations make the model more understandable.

Fifth, we used the {\BpForms} annotations to systematically identify missing proteoforms that could help the model better explain how the MAPK pathway transduces signals. Specifically, we used {\BpForms} to identify two missing combinations of the individual protein modifications represented by the model and four missing reactions that involve these species (\autoref{fig:kholodenko}B). These additional species and reactions could help the model better capture the kinetics of MAPKK and MAPKKK activation and deactivation and, in turn, better capture how the pathway transduces signals.

Next, we used the {\BpForms} annotations to identify opportunities to merge the Kholodenko model with models of other signaling cascades. Specifically, we searched BioModels for other models that represent similar proteoforms. This analysis identified several models that represent EGFR, PI3K, S6K, and the transcriptional outputs of the MAPK pathway that could be composed with the Kholodenko model. Furthermore, this combination of models enabled us to identify emergent combinations of proteoforms that are missing from the individual models (\autoref{fig:kholodenko}C).

Lastly, to identify opportunities to merge the Kholodenko model with a model of metabolism, we used the {\BpForms} annotations to systematically identify unbalanced reactions with missing metabolites. This analysis identified four missing species that, if added to the Kholodenko model, would make the model composable with models of metabolism (\autoref{fig:kholodenko}D).

\subsubsection{Synthetic biology: Systematic identification of design constraints}
A promising way to engineer cells is to combine naturally-occurring parts, such as genes that encode metabolic enzymes, in an accommodating host, such as \textit{E. coli}. However, there are numerous potential barriers to transforming parts into other cells. For example, parts that require post-translational modifications cannot be transformed into cells which cannot synthesize the modifications. Currently, it is difficult to identify such design constraints because we have limited tools to describe the dependencies of parts. Here, we illustrate how {\BpForms} can systematically identify potential flaws in the design of a novel strain of \textit{E. coli} due to missing post-translational modification machinery.

First, we used the PDB and {\BpForms} to identify all of the modifications that have been observed in \textit{E. coli}. Second, we used the PDB and {\BpForms} to identify modifications which have never been observed in \textit{E. coli} and the proteins which contain these modifications. For example, we found that proteins that contain 4-hydroxproline (PDB CCD: \href{https://www.rcsb.org/ligand/HYP}{HYP}), such as collagen (UniProt: \href{https://www.uniprot.org/uniprot/P02452}{P02452}), potentially cannot be transformed into \textit{E. coli}. Third, we used the literature to confirm the absence of these modifications from \textit{E. coli} \cite{pinkas2011tunable, an2014engineered, yi2014biosynthesis}. \textcolor{attr}{Table~S5} lists the most common modifications which could constrain the transformation of proteins into \textit{E. coli}.

Bioengineers could use this information to more reliably modify strains by limiting designs to post-translationally compatible proteins or by co-transforming parts with their requisite post-translational modification machinery. Furthermore, the synthetic biology community could make such information more accessible for learning design rules by incorporating this information into parts repositories such as SynBioHub \cite{mclaughlin2018synbiohub}. This information would enable these repositories to function as dependency management systems for synthetic organisms, analogous to the Advanced Package Tool (APT) for Ubuntu packages.

\section{Discussion}

\subsection{Community adoption as a common toolkit}
Realizing the full potential of {\BpForms} and {\BcForms} as formats for the primary structures of macromolecules will require acceptance by the omics, systems biology, and synthetic biology communities. We have begun to solicit users by submitting the {\BpForms} and {\BcForms} grammars to the FAIRsharing registry of standards\dontcite{sansone2019fairsharing} and the EDAM ontology of formats\dontcite{ison2013edam}, contributing the alphabets of residues and the ontology of crosslinks to BioPortal\dontcite{salvadores2013bioportal}, proposing a protocol for using {\BpForms} with SBOL, and helping the PRO consortium use {\BpForms} to represent proteoforms. To further encourage community adoption, we plan to encourage the developers of central repositories of DNA, RNA, and protein modifications such as MethSMRT \cite{ye2016methsmrt}, the PDB, and RMBase \cite{xuan2017rmbase} to export their data in {\BpForms} format. We also plan to stimulate discussion among the BioPAX, CellML, and SBML communities about formalizing our integrations of {\BpForms} and {\BcForms} with their formats. Additionally, we also plan to use the grammars to generate parsers for other languages, such as C\textsuperscript{++}, to help developers incorporate {\BpForms} and {\BcForms} into software tools.


\subsection{Community adoption as standards}
Because {\BpForms} and {\BcForms} aim to help researchers exchange information, we believe that the alphabets of residues, the ontology of crosslinks, and the grammars should ultimately become community standards. To start, we encourage the community to contribute to {\BpForms} and {\BcForms} via Git pull requests. Going forward, we would like these resources to be governed by the community through an organization such as the Computational Modeling in Biology Network (COMBINE) \cite{hucka2015promoting}.

\subsection{Integrating closed chemical representations with open informatics representations to enable WC models}
{\BpForms} and {\BcForms} achieve abstract descriptions of macromolecules by combining a closed, defined grammar with open, extensible ontologies of residues and crosslinks. This hybrid approach enables {\BpForms} and {\BcForms} to integrate diverse data into chemically-concrete descriptions of a wide range of macromolecules. Achieving WC models swimilarly requires integrating heterogeneous data about a wide range of processes from a wide range of methods and sources into physically-concrete kinetic simulations. Consequently, we believe that hybrid open-closed approaches such as {\BpForms} and {\BcForms} will be essential for WC modeling. For example, we are developing a hybrid methodology that enables chemically-concrete coarse-grained simulations by using fine-grained reactions to describe the chemical semantics of coarse-grained reactions.

\subsection{Enabling multiscale models that bridge structural information with networks}
We have begun to use {\BpForms} and {\BcForms} to describe the chemical semantics of the species represented by network models. Going forward, we also plan to use {\BpForms} and {\BcForms} to help network models capture finer-grained mechanisms that involve combinatorial interactions, such as how methylation impacts transcription factor-DNA binding. To do this, we are developing a generalized rule-based modeling framework which encapsulates properties such as primary structures into species and links these properties to reactions and rate laws. We anticipate that this framework, together with {\BpForms} and {\BcForms}, will make it easier to build fine-grained kinetic models of complex processes such as transcriptional backtracking, ribosomal queuing, and tmRNA ribosomal rescuing and combine them into WC models.


\section{Conclusions}
The {\BpForms}-{\BcForms} toolkit abstracts the primary structure of polymers and complexes, including non-canonical residues, caps, crosslinks, nicks, and several types of missing information. Furthermore, the toolkit standardizes the representation of common residues and crosslinks while extensibly accommodating any residue and crosslink by supporting both centrally and user-defined abstractions of residues and crosslinks. The toolkit includes the most extensive alphabets of hundreds of DNA, RNA, and protein residues; the first ontology of common crosslinks; an intuitive coordinate system for the subunits, residues, and atoms in macromolecules; the first human and machine-readable grammar for composing residues, caps, crosslinks, and nicks into polymers and complexes; and user-friendly web, REST, command-line and Python interfaces. The toolkit is backward compatible with the IUPAC/IUBMB format to maximize compatibility with existing bioinformatics tools and knowledge. The toolkit can also be integrated with frameworks for network research such as BioPAX, CellML, SBML, and SBOL.

We anticipate that {\BpForms} and {\BcForms} will be valuable tools for omics, systems biology, and synthetic biology. First, the tools can help researchers precisely communicate information about macromolecules. For example, the tools can help experimentalists communicate observations of proteoforms and help bioinformaticians exchange information among databases of polymers and complexes. Similarly, the tools can make models and genetic designs more understandable by capturing the semantic meaning of the species represented by models and capturing the structures of the parts of synthetic organisms. For example, {\BpForms} could describe proteins produced by expanded genetic codes.

The tools can also help quality control information about macromolecules. For example, the tools could help researchers find errors in reconstructed proteoforms such as inconsistencies between the modified and translated sequences, merge duplicate entries in databases of proteoforms, and identify gaps and element imbalances in models.

In addition, {\BpForms} and {\BcForms} can help researchers integrate structural, epigenomic, transcriptomic, and proteomic information about macromolecules. For example, the tools can help researchers integrate observations of individual protein modifications into descriptions of entire proteoforms. The tools can also help researchers integrate databases of modified proteins into a model of post-translational processing, combine the model with models of other processes to create WC models, and refine the model by identifying missing combinations of protein states. Similarly, the tools can help bioengineers design biochemical networks by identifying parts that must be co-transformed with post-transcriptional and post-translational modification machinery. 


\section{Methods}
We designed {\BpForms} and {\BcForms} as separate, but interrelated tools, to provide users light-weight tools for the distinct use cases of describing polymers and complexes. We implemented the toolkit using Python, ChemAxon Marvin, Flask-RESTPlus, Lark, Open Babel \cite{o2011open}, YAML Ain't Markup Language, and Zurb Foundation. \textcolor{attr}{Additional File~1.10} provides more information about the implementation.

\section*{Declarations}
\subsection*{Availability of data and materials}
The web applications are located at \href{https://bpforms.org}{https:\allowbreak{}//\allowbreak{}bp\allowbreak{}forms.\allowbreak{}org} and \href{https://bcforms.org}{https:\allowbreak{}//\allowbreak{}bc\allowbreak{}forms.\allowbreak{}org}, the REST APIs are located at \href{https://bpforms.org/api}{https:\allowbreak{}//\allowbreak{}bp\allowbreak{}forms.\allowbreak{}org/\allowbreak{}api} and \href{https://bcforms.org/api}{https:\allowbreak{}//\allowbreak{}bc\allowbreak{}forms.\allowbreak{}org/\allowbreak{}api}, the command-line programs and Python libraries are available from PyPI, and the code and ontologies are available at \href{https://github.com/KarrLab}{https://\allowbreak{}git\allowbreak{}hub.\allowbreak{}com/\allowbreak{}Karr\allowbreak{}Lab}. 

{\BpForms} and {\BcForms} are available open-source under the MIT license. Optionally, a license for ChemAxon Marvin is needed to calculate protonation and tautomerization states and generate molecular visualizations. Free licenses are available for academic researchers.

{\BpForms} and {\BcForms} are platform independent. The installation of {\BpForms} and {\BcForms} requires Python 3.6 or higher, Open Babel, and, optionally, ChemAxon Marvin. A Docker image with these dependencies is available at \href{http://dockerhub.com/u/karrlab}{http://\allowbreak{}dock\allowbreak{}er\allowbreak{}hub.\allowbreak{}com/\allowbreak{}u/\allowbreak{}karr\allowbreak{}lab}.

Documentation, including installation instructions, is available at \href{https://docs.karrlab.org}{https://\allowbreak{}docs.\allowbreak{}karr\allowbreak{}lab.\allowbreak{}org}. Interactive Jupyter notebook tutorials are available at \href{https://sandbox.karrlab.org}{https://\allowbreak{}sand\allowbreak{}box.\allowbreak{}karr\allowbreak{}lab.\allowbreak{}org}.

This article refers to versions 0.0.9 of {\BpForms} and 0.0.2 of {\BcForms}.

\subsection*{Competing interests}
The authors declare that they have no competing interests.

\subsection*{Funding}
This work was supported by the National Institutes of Health [grant numbers R35 GM119771, P41 EB023912]; the National Science Foundation [grant number 1649014]; and the Engineering and Physical Sciences Research Council [grant number EP/L016494/1].

\subsection*{Authors' contributions}
PFL, YC, XZ, DAN, and JRK built the alphabets of residues and the ontology of crosslinks. 
XZ, BS, and JRK developed the software. 
XZ, DAN, and JRK developed the case studies.
PFL, YC, JAPC, and JRK wrote the manuscript.
All authors read and approved the final manuscript.

\subsection*{Acknowledgements}
We thank Chris Myers and Jacob Beal for helpful discussion about integrating {\BpForms} with SBOL and Nicola Hawes for help designing \autoref{fig:overview}.

\section*{References}
\printbibliography[heading=none]

\iftoggle{submission}{}{
    \FloatBarrier
}

\clearpage

\section*{Figure legends and boxes}



\iftoggle{submission}{}{\clearpage}

\iftoggle{submission}{}{\clearpage}

\iftoggle{submission}{}{\clearpage}
\begin{figure}[!ht]
\centering
\iftoggle{submission}{}{%
    \includegraphics[width=\textwidth]{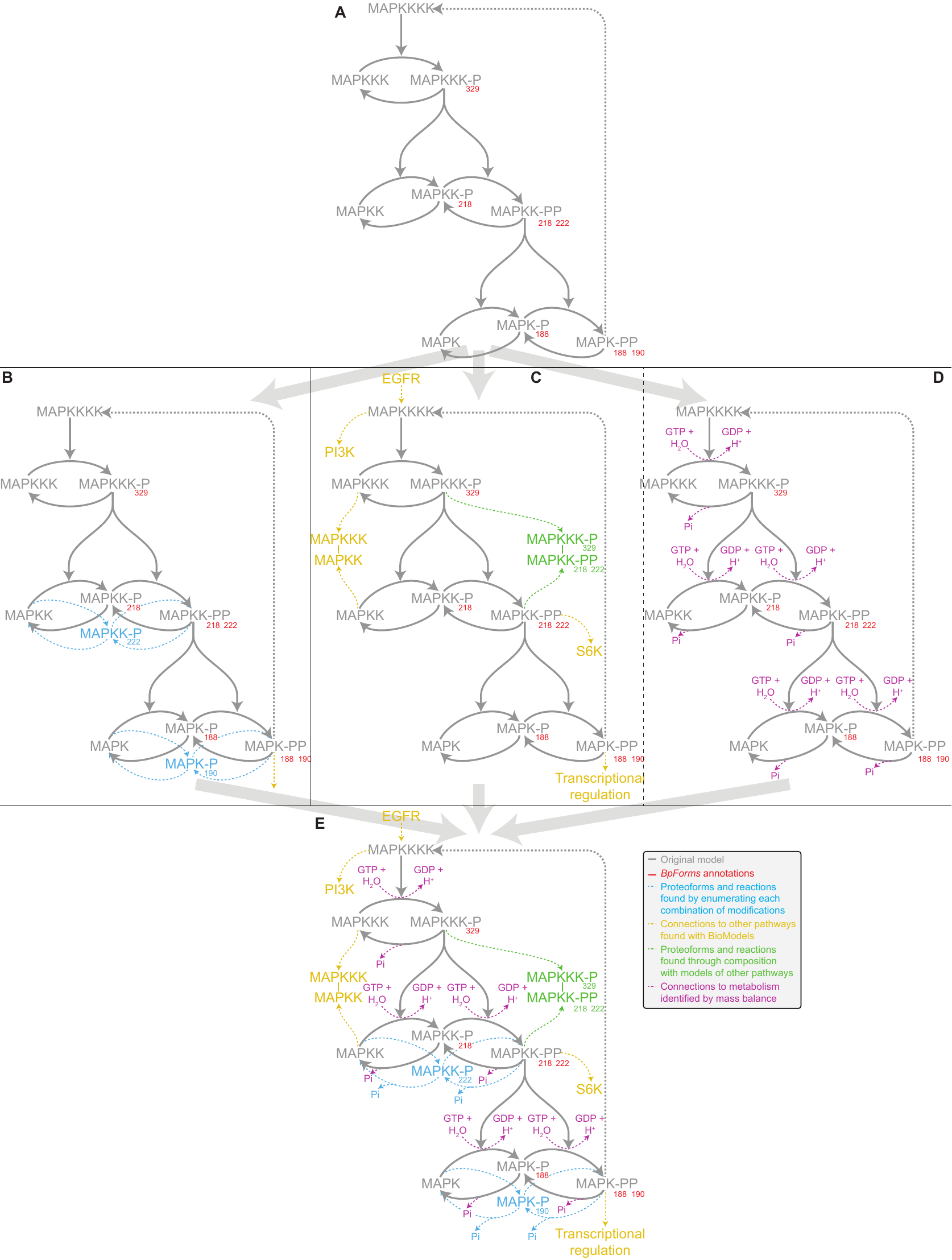}%
}%
\end{figure}
\iftoggle{submission}{}{\clearpage}
\begin{figure}[!ht]
  \caption{\label{fig:kholodenko} \textbf{{\BpForms} and {\BcForms} can facilitate the construction, expansion, composition, and refinement of fine-grained global intracellular networks.} For example, we used {\BpForms} to systematically identify ways to improve and expand the Kholodenko model of MAPK signaling (\textbf{A}, grey) by using {\BpForms} to capture the semantic meaning of each species (\textbf{A}, red), identify missing protein states (\textbf{B}, blue), identify other models that represent similar proteins which could be composed with the Kholodenko model (\textbf{C}, yellow) which could reveal additional missing combinations of species (\textbf{C}, green), and identify mass imbalances which indicate missing metabolites which could facilitate composition with metabolic models (\textbf{D}). Together, this could enable a substantially expanded model (\textbf{E}).}
\end{figure}

\clearpage
\begin{floatbox}[!ht]
\fbox{\begin{minipage}{\boxwidth}
\footnotesize
\setlength{\parskip}{8pt}

\textbf{Residue sequence}\\
This example illustrates how to use {\BpForms} to describe a DNA which begins with deoxyinosine.
\begin{tcolorbox}
\footnotesize\tt
\textcolor{bpforms}{\{\textcolor{val}{dI}\}}\textcolor{val}{ACGC}
\end{tcolorbox}

\textit{User-defined residues}\\
Residues which are not captured by our public alphabets can be captured within descriptions of polymers. This example illustrates how to describe a protein which ends with N\textsuperscript{5}-methyl-L-arginine.
\begin{tcolorbox}
\footnotesize\tt
\textcolor{val}{CRGN}{\color{bpforms}[\\
\hspace*{6.5ex}\textcolor{attr}{id}: "\textcolor{val}{AA0305}"\\
\hspace*{4ex}| \textcolor{attr}{structure}: "\textcolor{val}{OC(=O)[C@H](CCCN(C(=[NH2])N)C)[NH3+]}"\\
\hspace*{4ex}| \textcolor{attr}{l-bond-atom}: \textcolor{val}{N16-1}\\
\hspace*{4ex}| \textcolor{attr}{r-bond-atom}: \textcolor{val}{C2}\\
\hspace*{4ex}| \textcolor{attr}{l-displaced-atom}: \textcolor{val}{H16+1}\\
\hspace*{4ex}| \textcolor{attr}{l-displaced-atom}: \textcolor{val}{H16}\\
\hspace*{4ex}| \textcolor{attr}{r-displaced-atom}: \textcolor{val}{O1}\\
\hspace*{4ex}| \textcolor{attr}{r-displaced-atom}: \textcolor{val}{H1}\\
\hspace*{4ex}| \textcolor{attr}{name}: "\textcolor{val}{N5-methyl-L-arginine}"\\
\hspace*{4ex}| \textcolor{attr}{synonym}: "\textcolor{val}{delta-N-methylarginine}"\\
\hspace*{4ex}| \textcolor{attr}{synonym}: "\textcolor{val}{N5-carbamimidoyl-N5-methyl-L-ornithine}"\\
\hspace*{4ex}| \textcolor{attr}{identifier}: "\textcolor{val}{MOD:00310}" @ "\textcolor{val}{mod}"\\
\hspace*{4ex}| \textcolor{attr}{identifier}: "\textcolor{val}{CHEBI:21848}" @ "\textcolor{val}{chebi}"\\
\hspace*{4ex}| \textcolor{attr}{base-monomer}: "\textcolor{val}{R}"\\
\hspace*{4ex}| \textcolor{attr}{comments}: "\textcolor{val}{Generated by protein-arginine N5-methyltransferase (EC 2.1.1.-).}"\\
\hspace*{2ex}]
}
\end{tcolorbox}

\textbf{Crosslinks and nicks}\\
This example illustrates how to describe a peptide that contains a disulfide bond between the cysteines at the first and third positions and a nick between the cysteine and alanine at the first and second positions.
\begin{tcolorbox}
\footnotesize\tt
\textcolor{val}{C:AC} {\color{bpforms} | \textcolor{attr}{x-link}: [\\
\hspace*{6.5ex}\textcolor{attr}{id}: "\textcolor{val}{disulfide}"\\
\hspace*{4ex}| \textcolor{attr}{l}: \textcolor{val}{1} | \textcolor{attr}{r}: \textcolor{val}{3}\\
\hspace*{2ex}]
}\textbf{}
\end{tcolorbox}

\textit{User-defined crosslinks}\\
Crosslinks which are not captured by our public ontology can be described inline. This example illustrates how to describe a peptide that contains a disulfide bond between the cysteines at the first and third positions.
\begin{tcolorbox}
\footnotesize\tt
\textcolor{val}{CAC} {\color{bpforms} | \textcolor{attr}{x-link}: [\\
\hspace*{5.5ex} \textcolor{attr}{l-bond-atom}: \textcolor{val}{1S11} | \textcolor{attr}{r-bond-atom}: \textcolor{val}{3S11}\\
\hspace*{4ex}| \textcolor{attr}{l-displaced-atom}: \textcolor{val}{1H11} | \textcolor{attr}{r-displaced-atom}: \textcolor{val}{3H11}\\
\hspace*{4ex}| \textcolor{attr}{comments}: "\textcolor{val}{disulfide bond between 1C and 3C}"\\
\hspace*{2ex}]
}
\end{tcolorbox}

\textbf{Circularity}\\
This example illustrates how to describe a circular di-deoxyribonucleic acid.
\begin{tcolorbox}
\footnotesize\tt
\textcolor{val}{AC}
{\color{bpforms}
 | \textcolor{attr}{circular}
}
\end{tcolorbox}

\textit{Missing knowledge}\\
User-defined residues can also capture missing information about the mass, charge, location, and biosynthesis of residues. This example illustrates how to describe a protein which contains a methylated cysteine or asparagine at an unknown position between the fifth and tenth residues.
\begin{tcolorbox}
\footnotesize\tt
\textcolor{val}{CRGN}{\color{bpforms}[\\
\hspace*{7.0ex}\textcolor{attr}{base-monomer}: "\textcolor{val}{C}"\\
\hspace*{4ex}| \textcolor{attr}{delta-mass}: \textcolor{val}{12} | \textcolor{attr}{delta-charge}: \textcolor{val}{0}\\
\hspace*{4ex}| \textcolor{attr}{position}: \textcolor{val}{5}-\textcolor{val}{10} [\textcolor{val}{C}, \textcolor{val}{N}]\\
\hspace*{2ex}]\\}\textcolor{val}{EGYNNYCRAKYRGH}
\end{tcolorbox}

\end{minipage}}
\caption{\label{box:bpforms-grammar} Examples of the {\BpForms} grammar for describing polymers.}
\end{floatbox}

\clearpage
\begin{floatbox}[!ht]
\fbox{\begin{minipage}{\boxwidth}
\footnotesize
\setlength{\parskip}{8pt}

\textbf{Subunit composition}\\
This example illustrates how to use {\BcForms} to describe MalEFGK (Complex Portal: \href{https://www.ebi.ac.uk/complexportal/complex/CPX-1932}{CPX-1932}), a heteropentameric maltose ABC transporter.
\begin{tcolorbox}
\footnotesize\tt
\color{bpforms}
\textcolor{val}{MalE} + \textcolor{val}{MalF} + \textcolor{val}{MalG} + \textcolor{val}{2} * \textcolor{val}{MalK}
\end{tcolorbox}

\textbf{Crosslinks}\\
This example illustrates how to use the crosslinks ontology to describe a disulfide-linked antiparallel homodimer of disintegrin schistatin of \textit{Echis carinatus} (UniProt: \href{https://www.uniprot.org/uniprot/P83658}{P83658}).
\begin{tcolorbox}
\footnotesize\tt
\color{bpforms}
\textcolor{val}{2} * \textcolor{val}{P83658} \\
\hspace*{2ex} | \textcolor{attr}{x-link}: [\\
\hspace*{8.5ex} \textcolor{attr}{id}: "\textcolor{val}{disulfide}"\\
\hspace*{6ex} | \textcolor{attr}{l}: \textcolor{val}{P83658}(\textcolor{val}{1})-\textcolor{val}{7}\\
\hspace*{6ex} | \textcolor{attr}{r}: \textcolor{val}{P83658}(\textcolor{val}{2})-\textcolor{val}{12}\\
\hspace*{4ex} ]\\
\hspace*{2ex} | \textcolor{attr}{x-link}: [\\
\hspace*{8.5ex} \textcolor{attr}{id}: "\textcolor{val}{disulfide}"\\
\hspace*{6ex} | \textcolor{attr}{l}: \textcolor{val}{P83658}(\textcolor{val}{1})-\textcolor{val}{12}\\
\hspace*{6ex} | \textcolor{attr}{r}: \textcolor{val}{P83658}(\textcolor{val}{2})-\textcolor{val}{7}\\
\hspace*{4ex} ]
\end{tcolorbox}

\textit{User-defined crosslinks}\\
Crosslinks which are not captured by our public ontology can be defined within descriptions of complexes. This example illustrates how to describe the crosslinking of 10 kDa chaperonin (UniProt: \href{https://www.uniprot.org/uniprot/P9WPE5}{P9WPE5}) of \textit{Mycobacterium tuberculosis} with prokaryotic ubiquitin-like protein Pup (UniProt: \href{https://www.uniprot.org/uniprot/P9WHN5}{P9WHN5}) via a isoglutamyl lysine isopeptide bond (RESID: \href{https://annotation.dbi.udel.edu/cgi-bin/resid?id=AA0124}{AA0124}). Cells use this crosslink to mark 10 kDa chaperonin for proteasomal degradation.
\begin{tcolorbox}
\footnotesize\tt
\color{bpforms}

\textcolor{val}{P9WPE5} + \textcolor{val}{P9WHN5}\\
\hspace*{2ex}| \textcolor{attr}{x-link}:\\
\hspace*{4ex}[\\
\hspace*{8.6ex}\textcolor{attr}{l-bond-atom}: \textcolor{val}{P9WHN5}(\textcolor{val}{1})-\textcolor{val}{100N1-1}\\
\hspace*{6ex}| \textcolor{attr}{r-bond-atom}: \textcolor{val}{P9WPE5}(\textcolor{val}{1})-\textcolor{val}{63C2}\\
\hspace*{6ex}| \textcolor{attr}{l-displaced-atom}: \textcolor{val}{P9WHN5}(\textcolor{val}{1})-\textcolor{val}{100H1+1}\\
\hspace*{6ex}| \textcolor{attr}{l-displaced-atom}: \textcolor{val}{P9WHN5}(\textcolor{val}{1})-\textcolor{val}{100H1}\\
\hspace*{6ex}| \textcolor{attr}{r-displaced-atom}: \textcolor{val}{P9WPE5}(\textcolor{val}{1})-\textcolor{val}{63N1}\\
\hspace*{6ex}| \textcolor{attr}{r-displaced-atom}: \textcolor{val}{P9WPE5}(\textcolor{val}{1})-\textcolor{val}{63H1}\\
\hspace*{6ex}| \textcolor{attr}{r-displaced-atom}: \textcolor{val}{P9WPE5}(\textcolor{val}{1})-\textcolor{val}{63H1}\\
\hspace*{6ex}| \textcolor{attr}{comments}: "\textcolor{val}{isoglutamyl lysine isopeptide bond}"\\
\hspace*{4ex}]

\end{tcolorbox}

\end{minipage}}
\caption{\label{box:bcforms-grammar} Examples of the {\BcForms} grammar for describing complexes.}
\end{floatbox}

\end{document}